\documentstyle[art10]{article}

\begin{document}
\def\rf#1{(\ref{eq:#1})}
\def\lab#1{\label{eq:#1}}
\def\nonu{\nonumber}
\def\br{\begin{eqnarray}}
\def\er{\end{eqnarray}}
\def\be{\begin{equation}}
\def\ee{\end{equation}}
\def\eq{\!\!\!\! &=& \!\!\!\! }
\def\foot#1{\footnotemark\footnotetext{#1}}
\def\lb{\lbrack}
\def\rb{\rbrack}
\def\llangle{\left\langle}
\def\rrangle{\right\rangle}
\def\blangle{\Bigl\langle}
\def\brangle{\Bigr\rangle}
\def\llb{\left\lbrack}
\def\rrb{\right\rbrack}
\def\Blb{\Bigl\lbrack}
\def\Brb{\Bigr\rbrack}
\def\lcurl{\left\{}
\def\rcurl{\right\}}
\def\({\left(}
\def\){\right)}
\def\v{\vert}                     
\def\bv{\bigm\vert}               
\def\Bgv{\;\Bigg\vert}            
\def\bgv{\bigg\vert}              
\def\lskip{\vskip\baselineskip\vskip-\parskip\noindent}
\def\mskp{\par\vskip 0.3cm \par\noindent}
\def\sskp{\par\vskip 0.15cm \par\noindent}
\def\bc{\begin{center}}
\def\ec{\end{center}}
\def\Lbf#1{{\Large {\bf {#1}}}}
\def\lbf#1{{\large {\bf {#1}}}}

\def\tr{\mathop{\rm tr}}                  
\def\Tr{\mathop{\rm Tr}}                  
\newcommand\partder[2]{{{\partial {#1}}\over{\partial {#2}}}}
\newcommand\partderd[2]{{{\partial^2 {#1}}\over{{\partial {#2}}^2}}}
\newcommand\Bil[2]{\Bigl\langle {#1} \Bigg\vert {#2} \Bigr\rangle}  
\newcommand\bil[2]{\left\langle {#1} \bigg\vert {#2} \right\rangle} 
\newcommand\me[2]{\left\langle {#1}\right|\left. {#2} \right\rangle} 

\newcommand\sbr[2]{\left\lbrack\,{#1}\, ,\,{#2}\,\right\rbrack} 
\newcommand\Sbr[2]{\Bigl\lbrack\,{#1}\, ,\,{#2}\,\Bigr\rbrack} 
\newcommand\Gbr[2]{\Bigl\lbrack\,{#1}\, ,\,{#2}\,\Bigr\} } 
\newcommand\pbr[2]{\{\,{#1}\, ,\,{#2}\,\}}       
\newcommand\Pbr[2]{\Bigl\{ \,{#1}\, ,\,{#2}\,\Bigr\}}  
\newcommand\pbbr[2]{\lcurl\,{#1}\, ,\,{#2}\,\rcurl}  

\def\a{\alpha}
\def\b{\beta}
\def\c{\chi}
\def\d{\delta}
\def\D{\Delta}
\def\eps{\epsilon}
\def\vareps{\varepsilon}
\def\g{\gamma}
\def\G{\Gamma}
\def\grad{\nabla}
\def\h{{1\over 2}}
\def\k{\kappa}
\def\l{\lambda}
\def\L{\Lambda}
\def\m{\mu}
\def\n{\nu}
\def\o{\over}
\def\om{\omega}
\def\O{\Omega}
\def\p{\phi}
\def\P{\Phi}
\def\pa{\partial}
\def\tpa{{\tilde \partial}}
\def\bpa{{\bar \partial}}
\def\pr{\prime}
\def\ra{\rightarrow}
\def\lra{\longrightarrow}
\def\s{\sigma}
\def\S{\Sigma}
\def\t{\tau}
\def\th{\theta}
\def\Th{\Theta}
\def\z{\zeta}
\def\ti{\tilde}
\def\wti{\widetilde}
\newcommand\sumi[1]{\sum_{#1}^{\infty}}   
\newcommand\twomat[4]{\left(\begin{array}{cc}  
{#1} & {#2} \\ {#3} & {#4} \end{array} \right)}
\newcommand\threemat[9]{\left(\begin{array}{ccc}  
{#1} & {#2} & {#3}\\ {#4} & {#5} & {#6}\\
{#7} & {#8} & {#9} \end{array} \right)}
\newcommand\BDet[5]{\det_{{#1}}\left\Vert\begin{array}{cc}  
{#2} & {#3} \\ {#4} & {#5} \end{array} \right\Vert}   
\newcommand\Det[2]{\det_{{#1}} \left\Vert {#2} \right\Vert}
\newcommand\twocol[2]{\left(\begin{array}{cc}  
{#1} \\ {#2} \end{array} \right)}
\def\cA{{\cal A}}
\def\cB{{\cal B}}
\def\cC{{\cal C}}
\def\cD{{\cal D}}
\def\cE{{\cal E}}
\def\cF{{\cal F}}
\def\cG{{\cal G}}
\def\cH{{\cal H}}
\def\cI{{\cal I}}
\def\cJ{{\cal J}}
\def\cK{{\cal K}}
\def\cL{{\cal L}}
\def\cM{{\cal M}}
\def\cN{{\cal N}}
\def\cO{{\cal O}}
\def\cP{{\cal P}}
\def\cQ{{\cal Q}}
\def\cR{{\cal R}}
\def\cS{{\cal S}}
\def\cT{{\cal T}}
\def\cU{{\cal U}}
\def\cV{{\cal V}}
\def\cX{{\cal X}}
\def\cW{{\cal W}}
\def\cY{{\cal Y}}
\def\cZ{{\cal Z}}

\def\mark{\noindent{\bf Remark.}\quad}
\def\prop{\noindent{\bf Proposition.}\quad}
\def\exam{\noindent{\bf Example.}\quad}
\newtheorem{definition}{Definition}[section]
\newtheorem{proposition}{Proposition}[section]
\newtheorem{theorem}{Theorem}[section]
\newtheorem{lemma}{Lemma}[section]
\newtheorem{corollary}{Corollary}[section]
\def\proof{\par{\it Proof}. \ignorespaces} \def\endproof{{$\Box$}\par}
\newenvironment{Proof}{\proof}{\endproof} 
\def\cKP{{\sf cKP}~}
\def\cKPrm{${\sf cKP}_{r,m}$~}
\def\scKP{{\sf scKP}}
\newcommand\Back{{B\"{a}cklund}~}
\newcommand\DB{{Darboux-B\"{a}cklund}~}
\def\BH{{Burgers-Hopf}~}
\def\tQ{{\widetilde Q}}
\def\tit{{\tilde t}}
\def\hQ{{\widehat Q}}
\def\hb{{\widehat b}}
\def\hR{{\widehat R}}
\def\htt{{\hat t}}
\def\tv{{\tilde v}}
\def\Res{{\rm Res}}
\def\bt{{\bar t}}
\def\pai{\partial^{-1}}
\def\bD{{\bar D}}
\def\bpai{{\bar \partial}^{-1}}
\def\bcL{{\bar {\cal L}}}
\def\bP{{\bar \Phi}}
\def\bPsi{{\bar \Psi}}

\newcommand{\nit}{\noindent}
\newcommand{\ct}[1]{\cite{#1}}
\newcommand{\bi}[1]{\bibitem{#1}}
\newcommand\PRL[3]{{\sl Phys. Rev. Lett.} {\bf#1} (#2) #3}
\newcommand\NPB[3]{{\sl Nucl. Phys.} {\bf B#1} (#2) #3}
\newcommand\NPBFS[4]{{\sl Nucl. Phys.} {\bf B#2} [FS#1] (#3) #4}
\newcommand\CMP[3]{{\sl Commun. Math. Phys.} {\bf #1} (#2) #3}
\newcommand\PRD[3]{{\sl Phys. Rev.} {\bf D#1} (#2) #3}
\newcommand\PLA[3]{{\sl Phys. Lett.} {\bf #1A} (#2) #3}
\newcommand\PLB[3]{{\sl Phys. Lett.} {\bf #1B} (#2) #3}
\newcommand\JMP[3]{{\sl J. Math. Phys.} {\bf #1} (#2) #3}
\newcommand\PTP[3]{{\sl Prog. Theor. Phys.} {\bf #1} (#2) #3}
\newcommand\SPTP[3]{{\sl Suppl. Prog. Theor. Phys.} {\bf #1} (#2) #3}
\newcommand\AoP[3]{{\sl Ann. of Phys.} {\bf #1} (#2) #3}
\newcommand\RMP[3]{{\sl Rev. Mod. Phys.} {\bf #1} (#2) #3}
\newcommand\PR[3]{{\sl Phys. Reports} {\bf #1} (#2) #3}
\newcommand\FAP[3]{{\sl Funkt. Anal. Prilozheniya} {\bf #1} (#2) #3}
\newcommand\FAaIA[3]{{\sl Functional Analysis and Its Application} {\bf #1}
(#2) #3}
\def\TAMS#1#2#3{{\sl Trans. Am. Math. Soc.} {\bf #1} (#2) #3}
\def\InvM#1#2#3{{\sl Invent. Math.} {\bf #1} (#2) #3}
\def\AdM#1#2#3{{\sl Advances in Math.} {\bf #1} (#2) #3}
\def\PNAS#1#2#3{{\sl Proc. Natl. Acad. Sci. USA} {\bf #1} (#2) #3}
\newcommand\LMP[3]{{\sl Letters in Math. Phys.} {\bf #1} (#2) #3}
\newcommand\IJMPA[3]{{\sl Int. J. Mod. Phys.} {\bf A#1} (#2) #3}
\newcommand\TMP[3]{{\sl Theor. Mat. Phys.} {\bf #1} (#2) #3}
\newcommand\JPA[3]{{\sl J. Physics} {\bf A#1} (#2) #3}
\newcommand\JSM[3]{{\sl J. Soviet Math.} {\bf #1} (#2) #3}
\newcommand\MPLA[3]{{\sl Mod. Phys. Lett.} {\bf A#1} (#2) #3}
\newcommand\JETP[3]{{\sl Sov. Phys. JETP} {\bf #1} (#2) #3}
\newcommand\JETPL[3]{{\sl  Sov. Phys. JETP Lett.} {\bf #1} (#2) #3}
\newcommand\PHSA[3]{{\sl Physica} {\bf A#1} (#2) #3}
\newcommand\PHSD[3]{{\sl Physica} {\bf D#1} (#2) #3}
\newcommand\JPSJ[3]{{\sl J. Phys. Soc. Jpn.} {\bf #1} (#2) #3}
\newcommand\JGP[3]{{\sl J. Geom. Phys.} {\bf #1} (#2) #3}

\setlength{\baselineskip}{2.6ex}

\title{From One-Component KP Hierarchy\\ to Two-Component KP Hierarchy 
and Back}

\author{H. Aratyn${}^1$, E. Nissimov${}^{2,3}$ and S. Pacheva${}^{2,3}$ \\
{\em ${}^1$ Department of Physics, University of Illinois at Chicago}\\
{\em 845 W. Taylor St., Chicago, IL 60607-7059, U.S.A.}\\
{\em ${}^2$ Institute of Nuclear Research and Nuclear Energy} \\
{\em Boul. Tsarigradsko Chausee 72, BG-1784 $\;$Sofia, Bulgaria}\\ 
{\em ${}^3$Department of Physics, Ben-Gurion University of the Negev }\\
{\em Box 653, IL-84105 $\;$Beer Sheva, Israel}}

\maketitle
\begin{abstract}
\setlength{\baselineskip}{2.6ex}   

We show that the system of the standard one-component KP hierarchy endowed 
with a special infinite set of abelian additional symmetries, generated by 
squared eigenfunction potentials, is equivalent to the two-component 
KP hierarchy.
  
\end{abstract}  
\setlength{\baselineskip}{2.6ex}

\section*{Background Information on the KP Hierarchy and Ghosts Symmetries.}

The starting point of our presentation is the pseudo-differential Lax 
operator $\cL$ obeying KP evolution equations  
w.r.t. the multi-time $(t) \equiv (t_1 \equiv x, t_2 ,\ldots )$ :
\be
\cL = D + \sum_{i=1}^{\infty} u_i D^{-i} \qquad ; \qquad
\partder{\cL}{t_l} = \Sbr{\(\cL^{l}\)_{+}}{\cL} \quad , \; \;
l = 1, 2, \ldots
\lab{lax-eq}
\ee
The symbol $D$ stands for the differential
operator $\pa/\pa x$, whereas $\pa \equiv \pa_x$ will denote derivative of a
function. 
Equivalently, one can represent Eq.\rf{lax-eq} in terms of the 
dressing operator $W$ whose pseudo-differential series are 
expressed in terms of the so called tau-function $\t (t)$ :
\be
\cL = W D W^{-1} \quad ,\quad \partder{W}{t_l} = - \(\cL^l\)_{-} W \quad ,
\quad W = \sum_{n=0}^{\infty} \frac{p_n \( - [\pa]\)\t (t)}{\t (t)} D^{-n}
\lab{W-main}
\ee
with the notation: $[y] \equiv \( y_1,  y_2/2 , y_3/3 ,\ldots \)$
for any
multi-variable $(y) \equiv \( y_1 ,y_2 ,y_3 ,\ldots \)$ and with $p_k (y)$
being the Schur polynomials.
In the present approach a basic notion is that of (adjoint) eigenfunctions 
$\P (t),\, \Psi (t)$ of the KP hierarchy satisfying :
\be
\partder{\Phi}{t_k} = \cL^{k}_{+}\bigl( \Phi\bigr) \qquad; \qquad
\partder{\Psi}{t_k} = - \(\cL^{*} \)^{k}_{+}\bigl( \Psi\bigr)
\lab{eigenlax}
\ee
The Baker-Akhiezer (BA) ``wave''
functions $\psi_{BA} (t,\l ) = W (\exp(\xi (t,\l )))$
and its adjoint $\psi_{BA}^{*} (t,\l )= (W^{*})^{-1}(\exp(-\xi (t,\l )))$ 
(with $\xi (t,\l ) \equiv \sum_{l=1}^\infty t_l \l^l$)
are (adjoint) eigenfunctions satisfying additionally the spectral equations 
$\cL^{(*)}\bigl( \psi_{BA}^{(*)} (t,\l )\bigr) =  \l \psi_{BA}^{(*)} (t,\l )$.

Throughout this note we will rely on an important tool provided by the
spectral representation of eigenfunctions \ct{ridge}.
The spectral representation is equivalent to the following statement.
$\P$ and $\Psi$ are (adjoint) eigenfunctions if and only if they obey the
integral representation:
\br
\P (t) &=&  \int dz\, {e^{\xi (t-t^{\pr} ,z )}\o z} 
{ \t (t - [z^{-1}]) \t (t^{\pr} + [z^{-1}]) \o \t (t)  \t (t^{\pr} )  } \, 
  \P \( t^{\pr} + [z^{-1}]\) 
\lab{spec1t-a}\\
\Psi (t) &=& \int d z\, {e^{\xi (t^{\pr}-t ,z )}\o z} 
{ \t (t + [z^{-1}]) \t (t^{\pr} - [z^{-1}]) \o \t (t)  \t (t^{\pr} )  } \,
\Psi (t^{\pr} - [z^{-1}])
\lab{spec2t-a}
\er
where $\int dz$ denotes contour integral around origin.

One needs to point out that the proper understanding of 
Eqs.\rf{spec1t-a} and \rf{spec2t-a} (as in the case of original Hirota 
bilinear identities)
requires, following \ct{ldickey}, expanding of the integrand in 
\rf{spec1t-a} and \rf{spec2t-a} 
as formal power series w.r.t. in $t^{\pr}_n -t_n$, $n=1,2,\ldots $.


Consider now an infinite system of independent (adjoint) eigenfunctions
$\lcurl \P_j ,\Psi_j\rcurl_{j=1}^{\infty}$ of the standard KP hierarchy
Lax operator $\cL$ and define the following infinite set of the 
additional ``ghost'' symmetry flows \ct{hungry}:
\br
\partder{}{\bt_s} \cL &=& \Sbr{\cM_s}{\cL} \qquad ,\qquad
\cM_s = \sum_{j=1}^s \P_{s-j+1} D^{-1} \Psi_j 
\lab{ghost-s}  \\
\partder{}{\bt_s} \P_k \eq \sum_{j=1}^s \P_{s-j+1} S_{k,j} \; -\;
\P_{k+s} \quad ; \quad
\partder{}{\bt_s} \Psi_k = \sum_{j=1}^s \Psi_j S_{s-j+1,k}
\; + \; \Psi_{k+s} 
\lab{M-s-eigenf} \\
\partder{}{\bt_s} F &=& \sum_{j=1}^s \P_{s-j+1} S \(F \Psi_j \) \quad ; \quad
\partder{}{\bt_s} F^\ast = \sum_{j=1}^s \Psi_j S \( \P_{s-j+1} F^\ast\)
\lab{M-s-generic}
\er
where $s,k =1,2,\ldots$ , and $F$ ($F^{*}$) denote generic (adjoint) 
eigenfunctions which do not belong to the ``ghost''
symmetry generating set $\lcurl \P_j ,\Psi_j\rcurl_{j=1}^{\infty}$.
Moreover, we used abbrevations 
$S_{k,j} \equiv S \( \P_k , \Psi_j \)= \pai \( \P_k  \Psi_j \)$ to denote
the 
so called squared eigenfunction potentials (SEP) \ct{oevel,ridge} for which we 
find the  ``ghost'' symmetry flows :
\be
\partder{}{\bt_s} S_{k,l} = S_{k,l+s} - S_{k+s,l} + \sum_{j=1}^s S_{k,j}
S_{s-j+1,l}
\lab{ghost-S}
\ee
Eqs.\rf{M-s-eigenf} become for the first ``ghost'' symmetry flow
$ \bpa \equiv \pa/ \pa {\bt_1} $:
\be
\bpa \P_k = \P_1 S_{k,1} - P_{k+1} \quad , \quad
\bpa \Psi_k = \Psi_1 S_{1,k} + \Psi_{k+1} 
\quad ; \quad \bpa F = \P_1 \pai \(\Psi_1 F\) 
\lab{ghost-1}
\ee

It is easy to show that the ``ghost'' symmetry flows
$\pa / \pa {\bt_s}$ from Eqs.\rf{ghost-s}-\rf{M-s-generic}
commute. This can be done by proving that the $\pa$-pseudo-differential
operators $\cM_s $ \rf{ghost-s} satisfy the zero-curvature equations
$ \pa \cM_r / \pa {\bt_s} - \pa \cM_s /\pa \bt_r - \sbr{\cM_s}{\cM_r} = 0$.

\section*{Lax Representation for the Ghosts Flows}

We now show that the ``ghost'' symmetry flows from Eqs. 
\rf{ghost-s}-\rf{M-s-generic} admit their own Lax representation
in terms of the pseudo-differential Lax operator $\bcL$ w.r.t. 
multi-time $({\bar t}) \equiv ({\bar t}_1 \equiv 
{\bar x}, {\bar t}_2 ,\ldots )$.
While showing it we will make contact with the structure defining 
the affine coordinates on the Universal Grassmannian Manifold (UGM) 
\ct{takasaki,magrietal}. For this purpose we define objects:
\be
{\bar w}_i = { \P_{i+1} \o \P_{1} } \quad ; \quad i=0,1,2, {\ldots} 
\lab{bar-dressi}
\ee
which can be grouped together into the Laurent series expansion:
\be
{\bar w} = \sumi{i=0} { \P_{i+1} \o \P_{1} } z^{-i} =
1+ \sumi{l=1} { \P_{l+1} \o \P_{1} } z^{-l}
\lab{bar-dress}
\ee
{}From Eq.\rf{M-s-eigenf} we find that the action of the ``ghost'' 
symmetry flows on ${\bar w}_i$ takes a form:
\be 
\partder{}{\bt_s} {\bar w}_k  = - {\bar w}_{k+s} +
\sum_{l=0}^s {\bar w}_l {\overline W}^{(s-l)}_{k} 
\lab{t-bar-dress}
\ee
where the coefficients $ {\overline W}^{(j)}_{k}$ are given in terms of 
$SEP$-functions as
\br
{\overline W}^{(j)}_{k} &=& S_{k+1,j} - {\bar w}_k S_{1,j} 
\quad ; \quad j =1,2, {\ldots}  , k =0,1,2, {\ldots} 
\lab{cap-bar-W}\\
{\overline W}^{(0)}_{k} &=& {\bar w}_k = { \P_{k+1} \o \P_{1} } \; .
\lab{cap-bar-W0}
\er
They in turn satisfy the following flow equations resulting from those 
in \rf{ghost-S} :
\be
\partder{}{\bt_s} {\overline W}_k^{(j)}  =  {\overline W}^{(j+s)}_k -  
{\overline W}^{(j)}_{k+s}  +
\sum_{l=1}^s {\overline W}^{(s-l)}_{k} {\overline W}^{(j)}_{l} 
\lab{t-bar-Wdress}
\ee
which provide an example of the matrix Riccati equations (see 
{\sl e.g.} \ct{magrietal}).
The coefficients ${\overline W}^{(j)}_{k}$ span the Laurent series :
\be
{\overline W}^{(j)} = z^j + \sumi{l=1} {\overline W}^{(j)}_{l} z^{-l}
\quad ; \quad j =0,1,2, {\ldots} 
\lab{barW-Laurent}
\ee
whose structure is reminiscent of the Laurent series defining the Sato
Grassmannian. The connection to the usual KP setup can now be established 
as follows.
We first introduce the well-known notion of a long derivative :
\be
{\overline \nabla}_s = \partder{}{\bt_s} + z^s =
e^{- \xi  ({\bar t} , z)} \partder{}{\bt_s} e^{\xi  ({\bar t} , z)}
\lab{long-der}
\ee
which together with Eq.\rf{barW-Laurent} allow us to cast both 
Eqs.\rf{t-bar-Wdress} and \rf{t-bar-dress} in a more compact form:
\br
{\overline \nabla}_s  {\bar w} &=& \sum_{l=0}^s {\bar w}_{l} {\overline W}^{(s-l)}
\lab{t-bar-LWdress0} \\
{\overline \nabla}_s  {\overline W}^{(j)}  &= & {\overline W}^{(j+s)} +
\sum_{l=1}^s {\overline W}^{(j)}_{l} {\overline W}^{(s-l)} 
\quad ; \quad j =1,2, {\ldots} 
\lab{t-bar-LWdress}
\er
{}From Eq.\rf{t-bar-LWdress} we obtain reccursive expressions for
${\overline W}^{(j)}$ with $j >0$ in terms of non-negative powers of ${\overline \nabla}_1$ acting on
${\bar w}$. Indeed, from \rf{t-bar-LWdress} with $j=0$ one finds
${\overline W}^{(1)} = {\overline \nabla}_1  {\bar w} -  {\bar w}_1 {\bar w}$ and so on.
Finally, by increasing $j$ one arrives at expansion
$ {\overline W}^{(j)} = \sum_{l=0}^j v^{(j)}_l {\overline \nabla}_1^l {\bar w}$,
which allows to rewrite Eq.\rf{t-bar-LWdress0} as
${\overline \nabla}_s  {\bar w} = 
\sum_{l=0}^s U^{(s)}_l {\overline \nabla}_1^l {\bar w}$ with some coefficients
$U^{(s)}_l$.
Using Eq.\rf{long-der} we obtain the standard evolution equation
$\pa {\overline \psi}_{BA} ({\bar t} , z) /\pa  {\bt_s} =
{\overline B}_s {\overline \psi}_{BA} ({\bar t} , z) $
with ${\overline B}_s = \sum_{l=0}^s U^{(s)}_l {\bar D}^l$ and 
the wave-function :
\be
{\overline \psi}_{BA} ({\bar t} , z) =   {\bar w} e^{\xi  ({\bar t} , z)} =
{\overline {\cal W}} e^{\xi  ({\bar t} , z)} \;\; ;\;\;
{\overline {\cal W}} \equiv 1+ \sumi{l=1} { \P_{l+1} \o \P_{1} } {\bar D}^{-l}
\lab{bar-dressope}
\ee
with $ {\bar D} \equiv \pa/ \pa \bt_1 $.
The evolution operators ${\overline B}_s$ satisfy
$\pa {\overline {\cal W}} / \pa  {\bt_s} = {\overline B}_s  {\overline {\cal W}} -
{\overline {\cal W}} {\bar D}^{s}$
and are, therefore, reproduced by the usual relation
$ {\overline B}_s = \( {\overline {\cal W}} {\bar D}^{s} {\overline {\cal W}}^{-1} \)_{+}$.

The standard KP Lax operator construction follows now upon defining
the $\bpa$-Lax operator
$ {\bcL } \equiv  {\overline {\cal W}} {\bar D} {\overline {\cal W}}^{-1} =
{\bar D} + \sum_{i=1}^\infty {\bar u}_i {\bar D}^{-i}$,
which enters the hierarchy equations
$\pa {\bcL } / \pa {\bt_s} = \sbr{{\bcL }^s_{+}}{{\bcL }}$ with
$ {\overline B}_s = {\bcL }^s_{+}$.
In this way we arrive at a new integrable system defined in terms of two
Lax operators $\cL$ and $\bcL$ with two different sets of evolution parameters
$t$ and ${\bar t}$ which we will call {\em double KP} system.
The double KP system can be viewed as ordinary one-component KP hierarchy
Eq.\rf{W-main} supplemented by infinite-dimensional additional symmetry 
structure given by Eqs.\rf{ghost-s}-\rf{M-s-generic}.

Let ${\bar \t} (t,\bt )$ be a tau-function associated with the
$\bpa$-Lax operator $\bcL$, then the following results follow from the above
discussion:
\be  
{\bar \t}(t,\bt ) = \P_1 (t,\bt ) \,\t (t,\bt ) \qquad ,\quad 
\frac{p_s \( -[\bpa]\) {\bar \t}}{{\bar \t}} = \frac{\P_{s+1}}{\P_1}
\;\; ; \;\; s=0,1,2, {\ldots} 
\lab{bar-tau}
\ee
where the $\t (t,\bt ) $ is tau-function of the original $\pa$-Lax 
operator $\cL$.
Moreover, for any generic eigenfunction $F$ of $\cL\,$, which does not
belong to the set $\lcurl \P_j \rcurl$ in \rf{ghost-s} and has ``ghost'' 
symmetry flows given by Eq.\rf{M-s-generic}, 
the function ${\bar F} \equiv F/\P_1$ is
automatically an eigenfunction of the ``ghost'' Lax operator ${\bar \cL}$ :
\be
\partder{}{\bt_s} \( F/\P_1\) = {\bcL }^s_{+} \( F/\P_1\)
\lab{ghost-s-F-P1} 
\ee
We will also introduce the \DB (DB) transformations:
\be
\bcL(n+1) = \( {1\o {\P_1^{(n+1)}}} \bD^{-1} {\P_1^{(n+1)}}\)
\bcL(n) \( {1\o {\P_1^{(n+1)}}} \bD {\P_1^{(n+1)}}\)  
\lab{DB-bar-L} 
\ee
for the  ``ghost'' KP Lax operator which have an additional property
of commuting with the ``ghost'' symmetries \rf{ghost-s}.
In the Eq.\rf{DB-bar-L} the the DB ``site'' index $(n)$ parametrizes
the DB orbit. The convention we adopt is that the index $(n)$ labels 
the particular $\bpa$-Lax operator $\bcL$ constructed above.
In terms of the original isospectral flows the DB transformations
take a form :
\be
\cL (n+1) = \( \P^{(n)}_1 D {\P_1^{(n)}}^{\, -1} \)\, \cL (n) \, 
\(\P^{(n)}_1  D^{-1} {\P_1^{(n)}}^{\, -1}\)
\lab{DB-L} 
\ee
where $ \cL (n)$ is the original Lax operator underlying our construction.
In this setting the tau-function ${\bar \t}$ appears, according to
Eq.\rf{bar-tau}, to be nothing but the tau-function associated with the
the Lax operator $\cL (n+1)$ at the site $(n+1)$, namely 
$ {\bar \t} = \t (n+1)$.

We can now present results for the adjoint eigenfunctions $\Psi_i$
which parallel those in Eqs.\rf{bar-tau}-\rf{ghost-s-F-P1} for the
eigenfunctions $\Phi_i$.
Defining ${\bar w}_k^{\ast} = \Psi_{k+1} / \Psi_1$ we find that :
\be 
\partder{}{\bt_s} {\bar w}_k^{\ast}  =  {\bar w}_{k+s}^{\ast} +
\sum_{l=0}^s {\bar w}_l^{\ast} {{\overline W}^{\ast\, (s-l)}}_{k} 
\lab{t-bar-dress-ast}
\ee
with the coefficients 
\br
{{\overline W}^{\ast\, (j)}}_{k}&=&
S_{j, k+1} - {\bar w}_k^{\ast} S_{j,1}
\quad ; \quad j =1,2, {\ldots}  , k =0,1,2, {\ldots} 
\lab{cap-bar-W-ast}\\
{{\overline W}^{\ast\, (0)}}_{k} &=& {\bar w}_k^{\ast} = 
{ \Psi_{k+1} \o \Psi_{1} } \; .
\lab{cap-bar-W0-ast}
\er

Let ${\widehat \t} (t,\bt )$ be a tau-function associated with the
$\bpa$-Lax operator $\bcL (n-2)$ at the DB site $(n-2)$.
Then the following results can be shown:
\be  
{\widehat \t}(t,\bt ) = \Psi_1 (t,\bt ) \,\t (t,\bt ) \qquad ,\quad 
\frac{p_s \( [\bpa]\) {\widehat \t}}{{\widehat \t}} = \frac{\Psi_{s+1}}{\Psi_1}
\;\; ; \;\; s=0,1,2, {\ldots} 
\lab{bar-tau-psi}
\ee
where $\t (t,\bt )$ is the tau-function of the original $\pa$-Lax 
operator $\cL$ (at the DB site $(n)$) .
Moreover, for any generic adjoint eigenfunction $F^{\ast}$ of $\cL\,$, 
which does not belong to the set $\lcurl \Psi_j \rcurl$ in \rf{ghost-s} 
and satisfies, therefore, the ``ghost'' symmetry flows given by 
Eq.\rf{M-s-generic}) the function $F^{\ast}/\Psi_1$ is
an adjoint eigenfunction of the ``ghost'' Lax operator ${\bar \cL} (n-2)$ :
\be
\partder{}{\bt_s} \( F^{\ast}/\Psi_1\) = 
-\({\bcL^{\ast} (n-2)}\)^s_{+} \( F^{\ast}/\Psi_1\)
\lab{ghost-s-Fast-Ps1} 
\ee

Let us list two other important identities which relate the tau-function $\t$
to the $SEP$-functions (using notation of Eqs.\rf{ghost-s}-\rf{M-s-generic}) :
\be
\frac{p_j \( [\bpa]\) {\t}}{{ \t}} =  - S_{1,j} \quad; \quad 
\frac{p_j \(- [\bpa]\) {\t}}{{ \t}} =   S_{j,1} \quad; \quad j \geq 1
\lab{sjone}
\ee

\section*{Embedding of Double KP System into Two-Component KP Hierarchy}

The two-component KP hierarchy \ct{U-T} is given by three tau-functions
$\t_{11}, \t_{12}, \t_{21}$ depending on two sets of multi-time variables 
$t, {\bar t}$ and obeying the following Hirota bilinear identities:
\br
&&\int dz  {e^{\xi ({\bar t} - {\bar t}^{\pr}, z)}\o z^2 } \t_{12} 
(t, {\bar t} -[z^{-1}] ) \t_{21} (t^{\pr}, {\bar t}^{\pr}+ [z^{-1}] )
= \nonu \\
\eq \int dz  e^{\xi ( t - t^{\pr}, z)} \t_{11} (t - [z^{-1}], {\bar t}) 
\t_{11}  (t^{\pr}+ [z^{-1}], {\bar t}^{\pr}) 
\lab{HBI-11}\\ 
&&\int dz {e^{\xi ({\bar t} - {\bar t}^{\pr}, z)} \o z}
\t_{12} (t, {\bar t} - [z^{-1}] ) \t_{11} (t^{\pr}, {\bar t}^{\pr}+ [z^{-1}] )
= \nonu \\
\eq \int dz {  e^{\xi (t - t^{\pr}, z)} \o z } \t_{11} (t - [z^{-1}], {\bar t})
\t_{12}  (t^{\pr}+ [z^{-1}], {\bar t}^{\pr}) 
\lab{HBI-12}\\ 
&&\int dz {e^{\xi ({\bar t} - {\bar t}^{\pr}, z)} \o z}
\t_{11} (t, {\bar t} - [z^{-1}] ) \t_{21} (t^{\pr}, {\bar t}^{\pr}+ [z^{-1}] )
= \nonu \\
\eq \int dz {  e^{\xi (t - t^{\pr}, z)} \o z } \t_{21} (t - [z^{-1}], {\bar t}) 
\t_{11}  (t^{\pr}+ [z^{-1}], {\bar t}^{\pr}) 
\lab{HBI-21}\\ 
&&\int dz  e^{\xi ({\bar t} - {\bar t}^{\pr}, z)} \t_{11} (t, {\bar t} -
[z^{-1}] ) \t_{11} (t^{\pr}, {\bar t}^{\pr}+ [z^{-1}] ) = \nonu \\ 
\eq \int dz 
{  e^{\xi (t - t^{\pr}, z)} \o z^2 } \t_{21} (t - [z^{-1}], {\bar t}) 
\t_{12}  (t^{\pr}+ [z^{-1}], {\bar t}^{\pr})
\lab{HBI-22}    
\er
We will now show that the double KP system defined in the previous section 
in terms of the tau-functions $\t, {\bar \t}, {\widehat \t}$,
will satisfy the Hirota identities \rf{HBI-11}-\rf{HBI-22}
upon the identification :
\be
\t = \t_{11} \quad ; \quad {\bar \t} = \t_{12} \quad ; \quad
{\widehat \t} = \t_{21}
\lab{tau-ident}
\ee
and upon making the obvious identication for the multi-time variables
$t$ and ${\bar t}$.

As an example of our method we will derive Eq.\rf{HBI-12} using the
technique which employs the spectral representations 
\rf{spec1t-a}-\rf{spec2t-a}.
Let $F$ be a wave-function for the $\cL$ Lax operator:
\be
F = \psi_{BA} (t, {\bar t}, \l) = { \t (t - [\l^{-1}], {\bar t}) \o
\t (t, {\bar t})}  e^{\xi (t, \l)}
\lab{F-psi-L}
\ee
According to Eq.\rf{ghost-s-F-P1} :
\be
{F \o \P_1} =  { \t (t - [\l^{-1}], {\bar t}) \o
{\bar \t} (t, {\bar t})}  e^{\xi (t, \l)}
\lab{FP-psi-L}
\ee
is an eigenfunction for the Lax operator $\bcL$ w.r.t. the multi-time $\bt$
and in view of Eq.\rf{spec1t-a} admits the spectral representation :
\be
{F \o \P_1}  (t, {\bar t}) =
\int dz {e^{\xi ({\bar t} - {\bar t}^{\pr}, z)} \o z}\,
{ {\bar \t} (t, {\bar t} -[z^{-1}] ) \t (t, {\bar t}^{\pr}+ [z^{-1}] )
\o {\bar \t} (t, {\bar t} ) {\bar \t} (t, {\bar t}^{\pr}) }
\, F ( t, {\bar t}^{\pr}+ [z^{-1}] )
\lab{F-o-P1}
\ee
Substituting Eq.\rf{F-psi-L} into the r.h.s. of Eq.\rf{F-o-P1} and 
Eq.\rf{FP-psi-L} into the l.h.s. of Eq.\rf{F-o-P1} we obtain:
\be
{\bar \t} (t, {\bar t}^{\pr})\, \t (t - [\l^{-1}], {\bar t})
= \int dz {e^{\xi ({\bar t} - {\bar t}^{\pr}, z)} \o z}
\, {\bar \t} (t, {\bar t} -[z^{-1}] )
\, \t (t- [\l^{-1}], {\bar t}^{\pr}+ [z^{-1}] )
\lab{F-o-P1a}
\ee
Choose now ${\bar F}$ to be a wave-function for the Lax operator $\bcL$ :
\be
{\bar F} = {\overline  \psi}_{BA} (t, {\bar t}, \l) = 
{ {\bar \t} (t, {\bar t}- [\l^{-1}]) \o  {\bar \t} (t, {\bar t})}  
e^{\xi ({\bar t}, \l)}
\lab{bF-psi-L}
\ee
then the function :
\be
{\bar F} \, \P_1 =  { {\bar \t} (t, {\bar t}- [\l^{-1}]) \o
{ \t} (t, {\bar t})}  \; e^{\xi ({\bar t}, \l)}
\lab{bFP-psi-L}
\ee
is an eigenfunction for the Lax operator $\cL$ w.r.t. the multi-time $t$
and in view of Eq.\rf{spec1t-a} admits the spectral representation :
\be
{\bar F} \, \P_1  (t, {\bar t}) =
\int dz {e^{\xi ( t - t^{\pr}, z)} \o z}
\, { {\t} (t -[z^{-1}], {\bar t}) {\bar \t} (t^{\pr}+ [z^{-1}], {\bar t} )
\o { \t} (t, {\bar t} ) { \t} (t^{\pr}, {\bar t}) }
\, F ( t^{\pr}+ [z^{-1}], {\bar t} )
\lab{bF-o-P1}
\ee
Substituting Eq.\rf{bF-psi-L} into the r.h.s. of Eq.\rf{bF-o-P1} and 
Eq.\rf{bFP-psi-L} into the l.h.s. of Eq.\rf{F-o-P1} we obtain:
\be
{\bar \t} (t, {\bar t}- [\l^{-1}])  \, \t (t^{\pr}, {\bar t})
= \int dz {e^{\xi ({t} - t^{\pr}, z)} \o z}\,
{ \t} (t-[z^{-1}], {\bar t} ) \, 
{\bar \t} (t^{\pr}+ [z^{-1}], {\bar t}- [\l^{-1}] )
\lab{bF-o-P1a}
\ee
Subtracting Eq.\rf{bF-o-P1a} from Eq.\rf{F-o-P1a} indeed reproduces
Hirota identity \rf{HBI-12} with identifications $t^{\pr}= {t}- [\l^{-1}]$ 
and ${\bar t}^{\pr} = {\bar t}- [\l^{-1}]$.
Proofs of the remaining Hirota identities \rf{HBI-11}, \rf{HBI-21} and
\rf{HBI-22} follow along similar lines.

\section*{From Two-Component KP Hierarchy to Double KP Hierarchy}

In this section we are going to show that the two-component KP hierarchy,
with the two sets of multi-times $t,\bt$, can be regarded as ordinary 
one-component KP hierarchy w.r.t. to one of the multi-times, {\sl e.g.} $t$, 
supplemented by an infinite-dimensional abelian algebra of additional 
(``ghosts'') symmetries, such that the second multi-time $\bt$ plays the role
of ``ghost'' symmetry flow parameters.

We first observe, by putting $\bt=\bt^{\pr}$, $t=t^{\pr}$ in Hirota identities
\rf{HBI-11} and \rf{HBI-22}, that the tau-function $\t_{11} (t,\bt )$ defines 
two one-component KP hierarchies ${\rm KP}_{11}$ and 
${\overline {\rm KP}}_{11}$ w.r.t. the multi-time variables $t$ and $\bt$,
respectively. This is because in the limits $\bt=\bt^{\pr}$, $t=t^{\pr}$ 
Eqs.\rf{HBI-11}, \rf{HBI-22} reduce to the ordinary one-component
KP Hirota identities.

Let us define :
\be
\P_1 (t, {\bar t})  \equiv {  \t_{12} (t, {\bar t}) \o \t_{11} (t, {\bar t})}
\quad ; \quad
\Psi_1 (t, {\bar t})  \equiv {\t_{21} (t, {\bar t}) \o 
\t_{11} (t, {\bar t})}
\lab{def-P1-Psi1}
\ee
These functions have the following important properties.
$\P_1$ turns out to be simultaneously an eigenfunction of the ${\rm KP}_{11}$
hierarchy and an adjoint eigenfunction of the ${\overline {\rm KP}}_{11}$ 
hierarchy.
Similarly, $\Psi_1$ is simultaneously an adjoint eigenfunction of 
the ${\rm KP}_{11}$ hierarchy
and an eigenfunction of the ${\overline {\rm KP}}_{11}$ hierarchy.
The proof proceeds by showing that $\P_1$ and $\Psi_1$ satisfy the
corresponding spectral representations \rf{spec1t-a} and \rf{spec2t-a}
as a result of taking special limits in Hirota identities
\rf{HBI-12} and \rf{HBI-21}.

As a consequence of these last properties we conclude that
$\t_{12}= \P_1 \t_{11}$ and $\t_{21} = \Psi_1 \t_{11}$ are also tau-functions
of ${\rm KP}_{11}$  and ${\overline {\rm KP}}_{11}$ hierarchies since they
can be regarded as DB transformations of $\t_{11}$.

Next, define :
\be
\P_j (t, {\bar t})  \equiv {p_{j-1} \( -[\bpa]\)  \t_{12} (t, {\bar t}) 
\o \t_{11} (t, {\bar t})}  \quad ; \quad
\Psi_j (t, {\bar t})  \equiv {p_{j-1} \( [\bpa]\)  \t_{21} (t, {\bar t}) 
\o \t_{11} (t, {\bar t})} 
\quad ; \quad j \geq 1 
\lab{def-Pj}
\ee
It turns out that $\P_j , \Psi_j$ are (adjoint) eigenfunctions
of ${\rm KP}_{11}$. We present the proof for $\P_j$ which goes as follows.
Substitute ${\bar t}^{\pr} = {\bar t} - [\l^{-1}]$ in \rf{HBI-12}. 
Using identity $\int dz F (z) / z (1- z/ \l) = F_{(-)} (\l) $
(where the subscript $(-)$ indicates taking the non-positive-power part of the
corresponding Laurent series), the r.h.s. of \rf{HBI-12} becomes:
\be
\int dz { 1 \o z} {1 \o \l - z } 
\t_{12} (t, {\bar t} - [z^{-1}] ) \t_{11} 
(t^{\pr}, {\bar t} - [\l^{-1}] + [z^{-1}] )  =
\t_{12} (t, {\bar t} - [\l^{-1}] ) \t_{11} 
(t^{\pr}, {\bar t}  )
\lab{t12t11}
\ee
and the Hirota Eq.\rf{HBI-12} simplifies now to:
\be
\t_{12} (t, {\bar t} - [\l^{-1}] ) \t_{11} 
(t^{\pr}, {\bar t}  ) = 
\int dz {  e^{\xi (t - t^{\pr}, z)} \o z } \t_{11} (t - [z^{-1}], {\bar t})
\t_{12}  (t^{\pr}+ [z^{-1}], {\bar t} - [\l^{-1}])
\lab{HBI-12a}
\ee
Upon expanding Eq.\rf{HBI-12a} in $\l$ and keeping the term of 
the order $\l^{j}$ we obtain Eq.\rf{spec1t-a} for $ \P \to \P_{j+1}$, with 
$\P_{j+1}$ as defined in \rf{def-Pj}. Consequently, the functions $\P_{j+1}$ 
from Eq.\rf{def-Pj} are eigenfunctions of the ${\rm KP}_{11}$ hierarchy.
The proof for $\Psi_j$ goes along the same lines.

Define now functions $S_{1,j}$ as
$S_{1,j} \equiv  p_{j} \( [\bpa]\)  \t_{11}  / \t_{11}$
for $j \geq 1$ (cf. Eq.\rf{sjone}).
We can also define the functions $S_{k,j}$ for $k >1$ via:
\be 
S_{k+1,j} = {\overline W}^{(j)}_{k} + {\bar w}_k  
{p_{j} \( [\bpa]\)  \t_{11} \o \t_{11}}
\lab{S-from-W}
\ee
where  ${\overline W}^{(j)}_k$ and $ {\bar w}_j$ are the known affine coordinates 
of UGM for ${\overline {\rm KP}}_{11}$ hierarchy satisfying 
Eqs.\rf{t-bar-dress} and \rf{t-bar-Wdress}. Recall that the latter flow
equations can be regarded as an equivalent definition of 
${\overline {\rm KP}}_{11}$ hierarchy.


Now, one can show that upon substitution of definitions \rf{S-from-W} and
${\bar w}_j=\P_{j+1} / \P_1$ (following from Eq.\rf{def-Pj}) into
the ${\overline {\rm KP}}_{11}$ structure 
Eqs.\rf{t-bar-dress} and \rf{t-bar-Wdress}, the latter two
equations go over into Eqs.\rf{M-s-eigenf} and \rf{ghost-S}
for $\P_j$ and $S_{k,j}$, which define an infinite-dimensional additional
``ghost'' symmetry structure of ${\rm KP}_{11}$ hierarchy.

This last observation shows that the isospectral evolution parameters $\bt$ 
of the one-com\-po\-nent ${\overline {\rm KP}}_{11}$ hierarchy
indeed play the role of parameters
of an infinite-dimensional abelian algebra of additional ``ghosts''
symmetries of the one-component ${\rm KP}_{11}$ hierarchy.
\mskp

\section*{Conclusions and Outlook.} 

In the present paper we have shown that, given an ordinary one-component
KP hierarchy, we can always construct a two-component KP hierarchy, embedding
the original one, in the following way. We choose an infinite set of
(adjoint-)eigenfunctions of the one-component KP hierarchy (such a choice is
always possible due to our spectral representation theorem \ct{ridge}),
which we use to construct an infinite-dimensional abelian algebra of
additional symmetries. The one-component KP hierarchy equipped with such
additional symmetry structure turns out to be equivalent to the standard
two-component KP hierarchy.

It is an interesting question for further study whether the origin of
higher multi-component KP hierarchies can be similarly traced back to 
one-component KP hierarchy endowed with an appropriate infinite-dimensional 
abelian additional symmetry structure, generalizing the above construction for
two-component KP hierarchy.
\mskp

{\bf Acknowledgements.} 
The authors gratefully acknowledge support by the NSF grant {\sl INT-9724747}.

\bibliographystyle{unsrt}

\end{document}